\documentclass[12pt,a4paper]{article}

\usepackage[utf8]{inputenc}
\usepackage[T1]{fontenc}
\usepackage{amsmath,amssymb,amsfonts}
\usepackage{algorithmic}
\usepackage{graphicx}
\usepackage{textcomp}
\usepackage{xcolor}
\usepackage{braket}
\usepackage{lipsum}
\usepackage{cite}
\usepackage{array}
\usepackage{hyperref}
\usepackage{orcidlink}
\usepackage{stfloats}
\usepackage{comment}

\usepackage[margin=1in]{geometry}
\date{}
\begin{document}

\title{\textbf{Quantum Evolution Kernel 
for Node-Attributed Graphs}}

\author{
    \textbf{Manuel Padovani}, \textbf{Alessandra Di Pierro}, \textbf{Claudia Daffara} \\[0.8em]
    \small Department of Computer Science, University of Verona, Verona, Italy \\
    \small \texttt{\{manuel.padovani, alessandra.dipierro, claudia.daffara\}@univr.it}
}

\maketitle
\vspace{-0.5cm}
\begin{abstract}
We present a novel approach to embed node-attributed graphs onto neutral-atom quantum computers, and propose an application to digit recognition. Our approach extends PASQAL's quantum evolution kernel (QEK) by introducing a node-level feature encoding based on three-level quantum states. To this purpose, we present a prototype showing how the resulting embedding can be leveraged to extract correlation-based information for classification tasks. The entire quantum process is simulated using Pasqal's Pulser library,  which inevitably imposes constraints on the size of the experimental datasets. We compare our method with classical baselines to evaluate its potential. We show that, for the digit recognition task, our method achieves a higher mean accuracy than the more commonly used Jensen-Shannon distance-based methods by utilizing the Wasserstein distance.
\end{abstract}

\vspace{0.5em}

\section{Introduction}

Graph machine learning is a well-studied topic, given the large number of real-world problems that can be naturally represented as graphs. Classically, many kernels have been proposed to capture structural information, such as Weisfeiler-Lehman (WL) graph kernels \cite{shervashidze2011weisfeiler}, random walk kernels, shortest-path kernels, graphlet kernels, and many others \cite{kriegeSurveyGraphKernels2020}. However, these methods typically ignore continuous node features, which play an important role in node-attributed graphs \cite{maDeepLearningGraphs2021}, with applications ranging from molecular property prediction to handwritten character recognition.

To address this limitation, a major change occurred with the introduction of Graph Convolutional Networks (GCNs) \cite{kipfSemiSupervisedClassificationGraph2017a}, which generalize convolutional neural networks to graph data by naturally incorporating node-level features, encouraging deep learning techniques on graph-structured data, including handwritten character recognition \cite{ganCharactersGraphsInterpretable2023, ganStructureAwareHandwrittenTexta}.
However, GCNs are theoretically bounded by the 1-WL isomorphism test, limiting their expressiveness in distinguishing non-isomorphic graphs \cite{morrisWeisfeilerLemanGo}.
Furthermore, GCNs suffer from over-smoothing as the number of layers increases \cite{chenEnhancingGraphConvolutional2025}, and training becomes computationally expensive as the graph size grows. Thus, exploring different paradigms is of paramount interest. In particular, a growing attention for analyzing graph data is shown in quantum machine learning, a machine learning paradigm that leverages quantum mechanics principles \cite{biamonteQuantumMachineLearning2017}. 

Several comprehensive reviews have surveyed this topic \cite{yuQuantumGraphLearning2023, tangQuantumGraphComputing2022}, where both digital and analog approaches have been explored. Digital approaches typically leverage quantum neural networks (QNNs) implemented in gate-based quantum computers \cite{aiQuantumGraphNeural2024}, also incorporating node-level attributes \cite{chenHybridQuantumClassicalGraph2021}. Analog quantum platforms, particularly neutral-atom devices, have shown great potential for graph learning due to their natural mapping of graph structures onto atomic arrays \cite{henryQuantumEvolutionKernel2021, henrietQuantumComputingNeutral2020a}, especially in the context of quantum kernels. More recently, node-feature encoding strategies for analog quantum devices have also been investigated, including recent work by the Pasqal group \cite{djellabiAttributedgraphsKernelImplementation2026}.

Despite these recent advances, the incorporation of multiple continuous node attributes into analog quantum graph kernels remains a challenging and mostly unexplored problem. Nevertheless, they constitute a particularly effective approach for graph learning. 

In this work, the focus is on quantum kernel methods~\cite{Schuld_2019}, since they have shown promising performance in the low-data regime~\cite{havlicekSupervisedLearningQuantumenhanced2019} and are particularly well suited for an analog computing paradigm with neutral-atom quantum computers in the NISQ era. Our primary objective is to assess the architectural feasibility of multi-attribute continuous encoding on analog hardware, providing a baseline for future large-scale implementations.

\subsection{Contribution}
We propose a quantum graph kernel method, based on node-level feature encoding. This method is designed for Pasqal's Rydberg atom QPUs, whose flexible topology allows a natural mapping between graph nodes and physical atoms \cite{albrechtQuantumFeatureMaps2023}. Following the Quantum Evolution Kernel (QEK) framework \cite{henryQuantumEvolutionKernel2021}, the graph structure is encoded onto the atomic register via successive layers of quantum evolution, preserving topology through atom positioning and the Rydberg blockade mechanism. 
 
The main contribution of this paper is the separate encoding of node-level features via local Raman laser pulses. This extends the Pasqal's approach in \cite{djellabiAttributedgraphsKernelImplementation2026}, which only utilizes Rydberg laser pulses for encoding both the graph's structure and the node-level features. In addition to the Rydberg state, we exploit the two hyperfine ground states of each atom, leading to a three-level (qutrit) system per atom. The long coherence times of the hyperfine states make them suitable for encoding node attributes, enriching each local Hilbert space to represent both local and structural features. 

To define our kernel we use the Wasserstein distance on probability distributions, obtained by measuring correlation and occupation observables, instead of the most commonly used Jensen-Shannon metric, which we show to be less effective on the digit recognition task. 

We evaluate our approach on the MNIST dataset, comparing against classical graph machine learning baselines \cite{hamiltonGraphRepresentationLearning}.

\section{Method}
\subsection{Pulses and Evolution}
Before introducing the quantum evolution, we briefly describe the physics of neutral atom arrays as implemented in modern devices such as Pasqal's platform \cite{henrietQuantumComputingNeutral2020a}. A neutral atom used for quantum computation has three relevant states: two low-energy hyperfine ground states, $\ket g$ and $\ket h$, and a high-energy Rydberg state, $\ket{r}$, used to enable interactions between nearby atoms. Two atoms both in the Rydberg state experience a van der Waals repulsion scaling as $1/R^6$, $R$ being the interatomic distance. This generates the Rydberg blockade effect: below a threshold distance $R_b$, the doubly-excited state $\ket{rr}$ becomes energetically disfavored, which is used to create entanglement between nearby atoms. For an array of atoms, the interaction Hamiltonian is
$$
H_{\text{int}} = \sum_{i<j} \frac{C_6}{R_{ij}^6} \hat{n}_i \hat{n}_j,
$$
where $\hat{n}_i = \ket{r}_i\bra{r}$, $R_{ij}$ is the interatomic distance between atoms $i$ and $j$, and $C_6$ depends on the atomic species and Rydberg level. This allows the encoding of any unit-disk graph\footnote{A unit-disk graph is a geometric graph whose nodes represent points in a 2D plane, where an edge connects two nodes if and only if their Euclidean distance is below a fixed threshold} by positioning atoms such that connected nodes have $R_{ij} \leq R_b$ and unconnected nodes have $R_{ij} > R_b$, so that the system's evolution naturally encodes the graph structure.

Apart from the interaction term, we can act on the neutral atoms via two types of lasers: the Raman laser \cite{henrietQuantumComputingNeutral2020a}  which couples the two low-energy states, $\ket g$ and $\ket h$; the Rydberg laser which couples the low-energy state $\ket g$ with the Rydberg state $\ket r$. The general Hamiltonian applied by each laser is:
$$ H = -\frac{\delta}{2} \sigma_z + \frac{\Omega}{2} \left[ \cos(\phi)\sigma_x + \sin(\phi)\sigma_y \right] +H_{\text{int}}, $$
where there are three time-dependent parameters: the detuning ($\delta$), the Rabi frequency ($\Omega$), and the phase of the laser ($\phi$); while the $\sigma_i$ are the Pauli matrices. These parameters allow high flexibility in the possible transformations.

\subsection{Mapping Node-Feature Graphs to Neutral Atoms}
\label{MappingNodeAtoms}
Given a graph $G=(V,E)$ with node features organized as a matrix $X \in \mathbb{R}^{|V|\times d}$, where $d$ is the number of local features, we first encode the graph structure in the atoms' positions using $H_{\text{int}}$, as described above.

The second step is encoding the node features. Each node's feature vector is mapped to a state obtained by applying rotations to the qubit corresponding to the $\ket{g}, \ket{h}$ levels of the neutral atom associated with the node. Specifically, we use the laser parameters ($\delta$, $\Omega$, $\phi$) to implement rotations $R_X(\theta _1)$ and $R_Y(\theta _2)$ on the Bloch sphere, where $\theta_i$ are functions of the feature values. Since features may lie in arbitrary ranges, we first normalize them to an appropriate interval (e.g., $[0, 2\pi]$ or $[0, \pi]$) suitable for the rotation gates.

If edge features are present, these can be encoded into the interatomic distances $R_{ij}$ by linearly mapping a feature in $[a_{\min}, a_{\max}]$ to a distance in $[d_{\min}, d_{\max}]$, where $d_{\min}$ depends on the optical tweezers and $d_{\max} < R_b$, so that edge features modulate the interaction strength via $C_6/R_{ij}^6$. However, we do not evaluate the impact of this mechanism experimentally in this work.

In summary, graph structure is encoded via atom positions, node features via local laser rotations, and edge features via interatomic distances. 

Enabling interactions requires exciting the system to the Rydberg states via Rydberg lasers, and since Rydberg coherence times are short, evolution times must remain brief.

The final encoding of each graph is obtained as 
\begin{align}
\ket{\Psi(t)} &= U_{\text{int}} \, U_{\text{local}} \, \ket{g}^{\otimes |V|}, \quad \text{where} \nonumber \\
U_{\text{local}} &= \prod_{i=1}^{|V|} R^i_X(X_{i,1}) R^i_Y(X_{i,2}) \quad \text{and} \nonumber \\
U_{\text{int}} &= \mathcal{T} \exp\left(-i \int_0^t \left( H_{\text{int}} + H_{\text{Rydberg}}(\tau) \right) d\tau\right).
\label{eq:evolution_equation}
\end{align}
\noindent
The rotations $R^i_X$ and $R^i_Y$ in $U_{\text{local}}$ are applied to the $i$-th neutral atom and are generated by Raman lasers; $H_{\text{Rydberg}}$ operator in $U_{\text {int}}$ represents the action of the Rydberg laser with constant Rabi frequency and constant detuning over the evolution time $t$, and $\mathcal{T}$ is the time-ordering operator.\footnote{Global Rydberg pulse: $\Omega=4\pi$ rad/$\mu$s, $\delta=0.2$ rad/$\mu$s; Raman encoding pulses: Blackman Waveform, $t=800\ \text{n}s$.}
Since a single qubit encodes at most two real parameters, $d$ node features yield $d/2$ distinct quantum states per graph, each corresponding to a different feature pair: the first state encodes $(X_{i,1}, X_{i,2})$ for all nodes, the second $(X_{i,3}, X_{i,4})$, and so on.

\subsection{Distributions From Measurements}
After obtaining the final state for each graph, we extract information by sampling the state multiple times. \footnote{Given the small number of atoms and the observables considered in our experiments, 500 shots per circuit were sufficient for our purpose.} Because typical neutral-atom hardware cannot distinguish all three atomic states directly, we use ground-Rydberg basis measurements, assigning bit 0 to either low-energy state and bit 1 to the Rydberg state. 

Note that, since node features are encoded as population transferred between $\ket{g}$ and $\ket{h}$ while only the $\ket{g}$ component is coupled to $\ket{r}$ by the Rydberg laser, the local encoding effectively modulates the population available for Rydberg excitation: probability mass moved into $\ket{h}$ is protected from transitioning to $\ket{r}$, thus reducing the atom's contribution to the interaction-driven entanglement and the excitation readout.

Similarly to pooling in GCNs, this outcome is aggregated into a size-invariant, graph-level representation.

We construct two distribution types: (i) for a bit string of length $|V|$ with Hamming weight $n$, we consider $f = n/|V| \in [0,1]$ and estimate $p(f)$ as the fraction of sampled strings with that weight, binned into a discrete distribution; (ii) we consider correlations $\langle Z_i Z_j \rangle - \langle Z_i \rangle \langle Z_j \rangle$, as in \cite{djellabiAttributedgraphsKernelImplementation2026}, similarly binned and normalized. Both distributions are scale-invariant with respect to $|V|$. For each graph we thus obtain one excitation and one correlation distribution per encoded feature pair.

\subsection{Distribution-based Quantum Kernel}
\label{sec:kernel}

As mentioned before, we aim to define a kernel based on the final data representation, namely distributions. Given a space $S$, a kernel is a symmetric, non-negative valued function $K: S \times S \rightarrow \mathbb{R}^+$ that takes two objects and outputs a real number measuring their similarity.

In quantum machine learning, a quantum kernel is typically defined as the inner product $K(\ket{\phi}, \ket{\psi}) = |\langle \phi | \psi \rangle|^2$ between embedded states, maximized at 1 for identical states and minimized at 0 for orthogonal ones. This definition is not directly applicable here, since each graph is mapped onto a collection of states producing multiple distributions $\{P_i^G\}$; the kernel itself is computed classically from these distributions, with "quantum" referring to the fact that these originate from measurements on quantum states.

Several distance measures exist between probability distributions. Entropy-based distances such as the Jensen-Shannon distance (JS), a symmetrization of the Kullback-Leibler divergence whose square root is a true metric, are favored but cannot capture geometric differences. A family of distances that preserve geometric information are the Wasserstein metrics \cite{rubnerEarthMoversDistance, givens1984wasserstein}. The Wasserstein distance of order $p$ between two probability distributions $P, Q$ is defined as:
\begin{equation}
    W_p (P,Q) = \left(\inf_{\mu \in \Pi(P,Q)} \int d(x,y)^p \, d\mu(x,y)\right)^{\frac{1}{p}},
\end{equation}
where $\Pi(P,Q)$ is the set of joint distributions with marginals $P, Q$ \cite{givens1984wasserstein}. In our experiments, we will use $W_1$, also known as the Earth Mover's Distance, which measures the minimum work to transform $P$ into $Q$. Unlike JS, $W_1$ requires no binning, avoiding discretization losses, at the cost of higher time complexity ($O(m \log m)$ vs.\ $O(m)$ \cite{nguyenStreamingSlicedOptimal2026}), where $m$ is the number of shots. This distance has already been used for the definition of graph kernels in the classical setting \cite{zhangMotifEntropyGraph2023}.

For a given distance $D$, we can define a kernel as:
\begin{equation}
 K(P_k^{G_1}, P_k^{G_2}) = \exp\left(-\gamma D(P_k^{G_1}, P_k^{G_2})\right) 
\end{equation}
where $\gamma > 0$ is a scale parameter. However, this is not automatically a positive semi-definite (PSD) kernel for any distance. In our setting, since the distributions are one-dimensional, $W_1$ is known to be conditionally negative definite, which guarantees that the kernel is PSD.

Since each graph produces multiple distributions $\{P_k^G\}_{k=1}^{N_{\text{dist}}}$, we combine the individual kernels using a weighted sum:
\begin{equation}
K_{\text{sum}}(G_1, G_2) = \sum_{k=1}^{N_{\text{dist}}} \omega_k \exp\left(-\gamma D(P_k^{G_1}, P_k^{G_2})\right)
\label{eq:weighted_kernel_sum}
\end{equation}
where $\omega_k \geq 0$ are optimizable weights such that $\sum_k \omega_k = 1$, allowing the model to favor the most informative distributions; $P_k^{G_j}$ is the $k$-th distribution associated with the $j$-th graph, and $N_{\text{dist}} = d$ is the total number of distributions (equal to the number of node-level features $d$). Furthermore, we consider a second weighted combination:
\begin{equation}
    K_{\text{prod}}(G_1, G_2) = \exp\left(-\gamma \sum_{k=1}^{N_{\text{dist}}} \omega_k D(P_k^{G_1}, P_k^{G_2})\right)
\label{eq:weighted_kernel_prod}
\end{equation}
which sums distances before exponentiating, making it a stricter, more sensitive similarity measure.



The resulting kernel is used within a Support Vector Classifier (SVC) to find the margin-maximizing hyperplane, employing a standard one-vs-one strategy for multi-class classification.

For each binary classifier, we jointly optimize $\gamma$, the regularization term $C$, and the weights $\omega_k$ in \eqref{eq:weighted_kernel_sum} via nested cross-validation: an outer grid search over $\gamma$ (5 log-spaced values in $[0.1,100]$), with $\omega_k$ optimized inside via Bayesian optimization (Gaussian process surrogate, expected improvement, inner 3-fold CV selecting $C \in \{0.1, 1.0, 10.0\}$), and a final outer-loop selection of $C \in \{0.1, 1.0, 10.0, 100.0\}$. The triplet $(\gamma, \boldsymbol{\omega}, C)$ maximizing macro-averaged F1 is retained. Note  that the outer-loop $C$ is the one reported, while inner-loop values are used only while optimizing $\boldsymbol{\omega}$.

\subsection{Graph Embedding}
\label{graph_embedding}

The quantum process is carried out using Pasqal's Pulser library \cite{Silv_rio_2022}, a Python library based on QuTip to specifically simulate neutral-atom quantum computers. The virtual device used in this work is \texttt{pulser.DigitalAnalogDevice}, which supports both Raman and Rydberg lasers. Furthermore, preprocessing must respect device constraints, notably the 2D unit-disk topology and a minimum interatomic distance of $4\ \mu\text{m}$ (full constraints can be found in the Pasqal documentation). To embed the graph onto the quantum device, we find ideal atom positions using a standard force-directed spring model that balances attractive forces between connected nodes and repulsive forces among all pairs. Node positions are updated iteratively until an equilibrium configuration is reached. When edge features are present, the forces are calibrated so that the equilibrium target distance is close to $d_{ij}^{\text{target}}$, which is obtained, in the interval $[4.5,6]\mu m$, via the linear mapping explained in Section \ref{MappingNodeAtoms} . Otherwise, all connected pairs use a constant target distance $d_{\min}$.
The preprocessing ends by checking the quality of the transformation. We filter graphs based on three criteria: $|V| \leq 9$ (computational reasons), $\text{MSE} \leq 1.5$ between target and achieved interatomic distances over connected pairs, and maximum distance from center $\leq 50.0~\mu\text{m}$. Graphs that do not meet these criteria are discarded.

\subsection{Experiments}
We have prepared a dataset based on the MNIST dataset \cite{lecun2010mnist}, consisting of 28x28 pixel images of handwritten digits across all 10 digit classes, by transforming each image into a graph.\footnote{The complete codebase for reproducibility—including preprocessing thresholds, and classical baselines optimization—is publicly available at \url{https://github.com/ManuP1406/QUANTUM_OCR_PUBLIC}}
This procedure is partially inspired by the preprocessing used in \cite{ganCharactersGraphsInterpretable2023}, which applies a graph machine learning model to handwritten Chinese characters. 

\begin{figure}[htbp]
    \centering
    \includegraphics[width=0.85\linewidth]{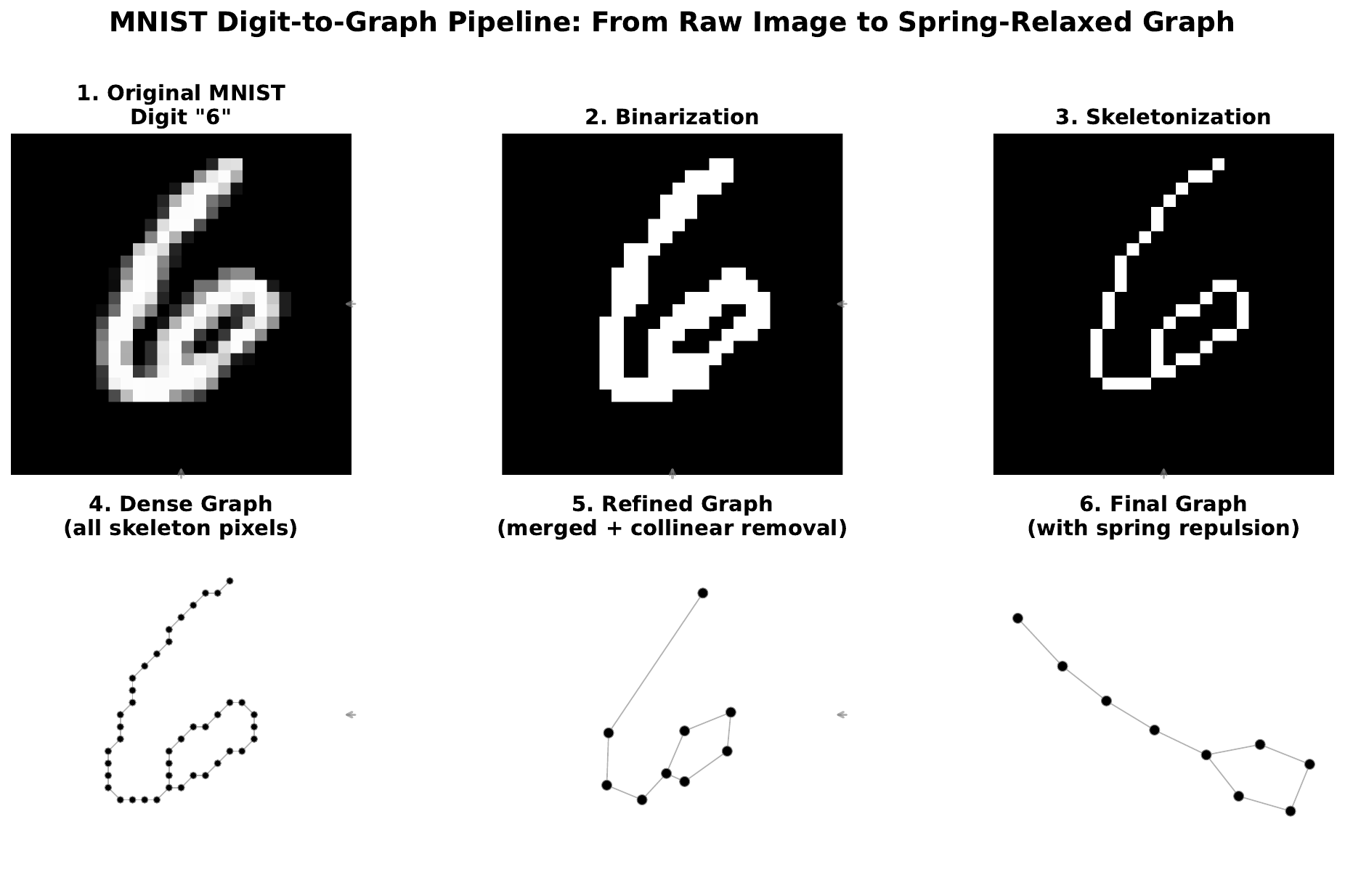}
    \caption{The picture shows the pipeline from the raw image to the final graph encodable onto the quantum device.}
    \label{fig:Image_pipeline}
\end{figure}

The process can be briefly explained as follows (Fig.~\ref{fig:Image_pipeline}):
\begin{itemize}
\item Binarization: all pixels are mapped to 0 (black pixel) or 1 (white pixel);
\item Skeletonization: the stroke width is reduced to a 1-pixel-wide stroke;
\item Dense graph: each pixel is associated with a node, and links between nearby pixels are formed;
\item Merge close nodes: nodes, which are at a distance (measured in pixels) below a given threshold, are merged together;
\item Remove collinear nodes: consecutive edges that form an almost-flat angle are merged by removing the intermediate node. This step further reduces the number of nodes, thus reducing the computational overhead.
\end{itemize}
\

After filtering, as described in section \ref{graph_embedding}, the retained portion of the dataset is approximately 80\% (48,018/60,000 training samples and 8,100/10,000 test samples). However, due to simulation constraints, only a small subset is used during training and evaluation. 

Regarding node-level features, we compute four values for each node: the original pixel coordinates $(x_{\text{pixel}}, y_{\text{pixel}})$ (linearly scaled to $[0, \pi/2]^2$ via a uniform rescaling that preserves the original aspect ratio), the deviation angle of consecutive edges, and the direction of the bisector of those edges. The deviation angle for a given pair of edges is defined as $|\pi - \theta|$, where $\theta$ is the angle between the two edge vectors. The direction is the angle of the bisector vector between the two edges, normalized to the range $[0, \pi)$.
For nodes with exactly two neighbors, we compute a single pair of features (deviation, direction) from the two incident edges. For nodes with three or four neighbors, we consider only the consecutive edge pairs with the largest angles: for three neighbors we take the single pair with the largest angle; for four neighbors we take the two pairs with the largest angles and sum their deviations and directions. For nodes with five or more neighbors, we take the three pairs with the largest angles. For nodes with zero or one neighbor, we set deviation to $0.9\pi$ and direction to $0$.

The restriction of the coordinates to $[0,\pi/2]$ ensures that only a partial population transfer from $\ket{g}$ to $\ket{h}$ occurs. If the population were completely moved to $\ket{h}$, that node would be disconnected from the Rydberg laser interaction (which only couples $\ket{g}$ to $\ket{r}$), potentially destroying the topology information encoded in the final state.

\begin{figure}[htbp]
    \centering
    \includegraphics[width=0.60\linewidth]{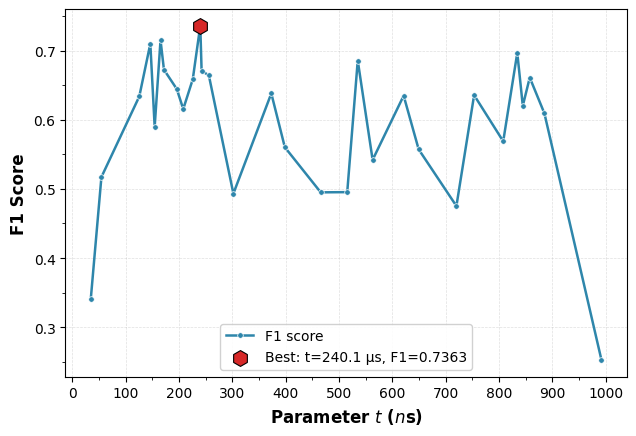}
    \caption{Bayesian optimization of the evolution parameter $t$ using the validation F1-score (80\% training, 20\% validation) on a 200-sample subset of the MNIST dataset. The optimal value is $t = 240.1\ n\text{s}$ (highlighted). For small $t$, the system does not generate sufficient entanglement, yielding low F1-score. For large $t$, long excitations cause oscillations between entangled states, leading to a sharp drop.}
    \label{fig:t_optimization} 
\end{figure}

The evolution time $t$ in (\ref{eq:evolution_equation}) is selected via Bayesian optimization using the macro-averaged F1-score as the objective. For each candidate $t$, we compute the quantum distributions, split the data into training (80\%) and validation (20\%) sets, train SVM classifiers, and evaluate the F1-score on the validation set. To keep the optimization feasible, this procedure is performed on a 200-sample subset of the MNIST training data. The optimization explores the range $t \in [25, 1000]~n\text{s}$ with 30 iterations and 5 initial random points. The value of $t$ that maximizes the validation F1-score is then selected for the final model. As shown in Fig.~\ref{fig:t_optimization}, the optimal value is $t = 240.1~n\text{s}$.

The quantum kernel model is compared to its version without node-level features to assess the contribution of local information, and we test different distances for the kernel (Wasserstein, Jensen-Shannon, Hellinger, and L2) to determine which one is more effective with our output distributions. For all metrics except Wasserstein, the occupation distributions are discretized into 10 bins (comparable to the typical number of nodes in our graphs, which is at most 9), while correlation distributions use 20 bins, as correlation values are concentrated near zero, especially for small evolution times, so finer binning helps increase the resolution. While all the quantum kernels are aggregated via the weighted sum in  \eqref{eq:weighted_kernel_sum}, for the Wasserstein metric we also implement the product aggregation in  \eqref{eq:weighted_kernel_prod} to test the influence of the aggregation strategy. 

We tested our method against classical baselines, such as Weisfeiler-Lehman (WL) graph kernels (implemented with GraKeL \cite{siglidis2020grakelgraphkernellibrary}). Note that these WL kernels use GraKeL's normalization, which reduces but may not fully eliminate graph-size bias, while our quantum distributions are size-invariant by construction. We evaluated different node label configurations (no features, positional only, angle only, combined), and a graph convolutional network (GCN). For the positional and angle configurations, node labels are obtained by discretizing the corresponding features into bins, while the combined configuration uses both. The GCN is implemented with PyTorch Geometric \cite{fey2019fastgraphrepresentationlearning} using two convolutional layers, batch normalization, and dropout to add regularization. Its input features are the same four node-level features used in our quantum kernel, but without the rescalings.

All models are evaluated using different training set sizes, with five random seeds for each size to assess robustness in the low-data regime, and on a fixed balanced test set of 15 samples per class, reporting accuracy as the primary metric.

\section{Results}

\subsection{Comparison of Distance Metrics}

\begin{figure}[htbp]
    \centering
    \includegraphics[width=0.8\linewidth]{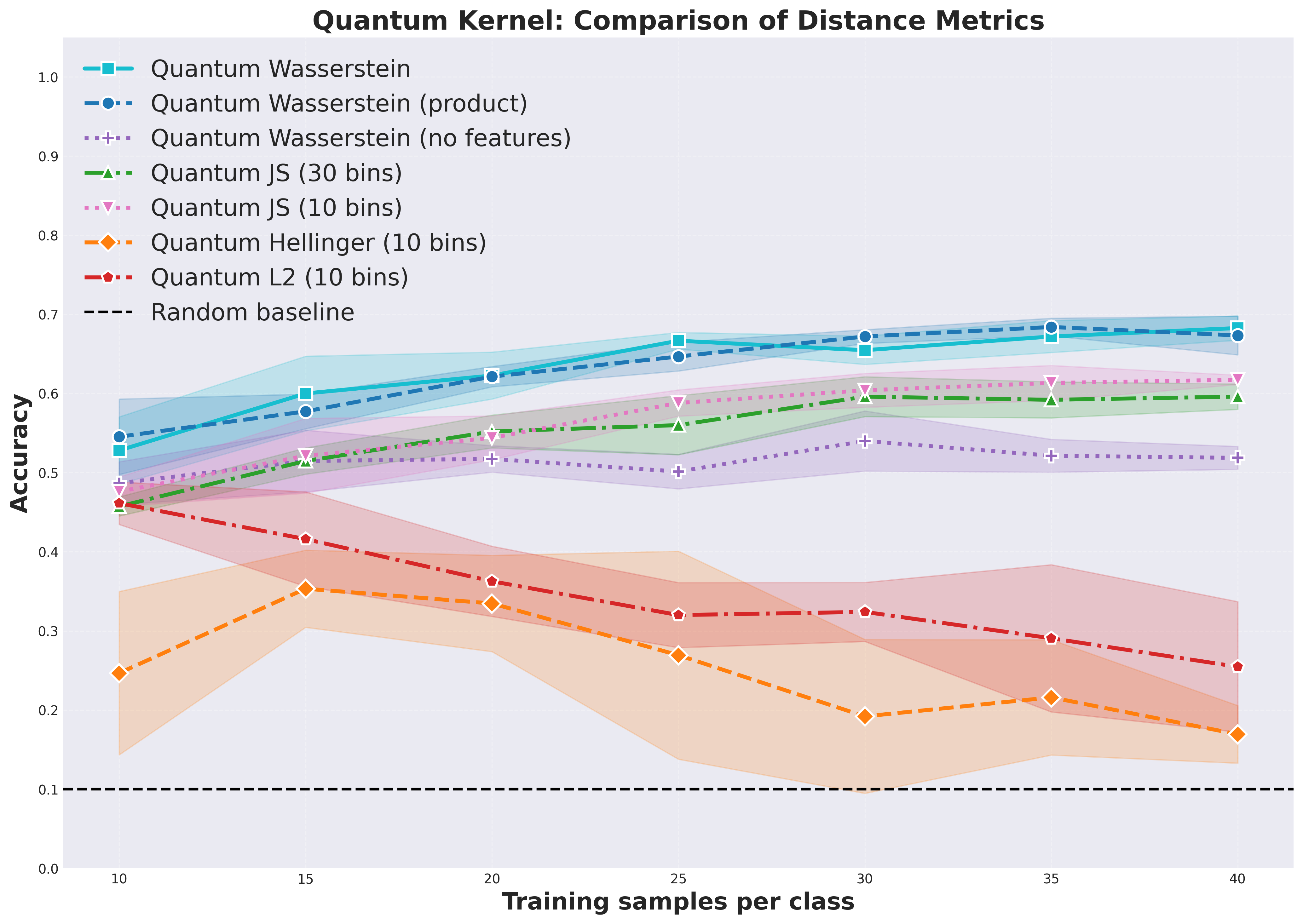}
    \caption{Comparison of the performances of different distance metrics. Wasserstein achieves the highest accuracy. Shaded regions indicate $\pm 1$ standard deviation over 5 seeds.}
    \label{fig:quantum_metrics}
\end{figure}

In Fig. \ref{fig:quantum_metrics} we show the accuracy as a function of the training samples per class for the quantum kernel using different distance metrics. The kernel based on Wasserstein distance achieves the highest mean accuracy for all training sizes, with both aggregation strategies producing comparable results; we therefore use the sum aggregation in the following comparison with classical baselines. The Jensen-Shannon distance performs slightly worse, and increasing the number of bins to 30 (way higher than the typical number of nodes in our graphs) leads to a drop in performance compared to the 10-bin version. In contrast, the Hellinger and L2 distances perform poorly, with accuracy even decreasing as the training set size increases, suggesting that these metrics are not suitable for capturing the separation between distributions in our setting. Finally, the Wasserstein kernel without node-level features leads to better performance than Hellinger and L2, but performs worse than all other cases that include local features. Thus, our simple feature encoding actually leads to better results, and seems promising for more complex applications.

\subsection{Comparison with Classical Baselines}

\begin{figure}[htbp]
    \centering
    \includegraphics[width=0.8\linewidth]{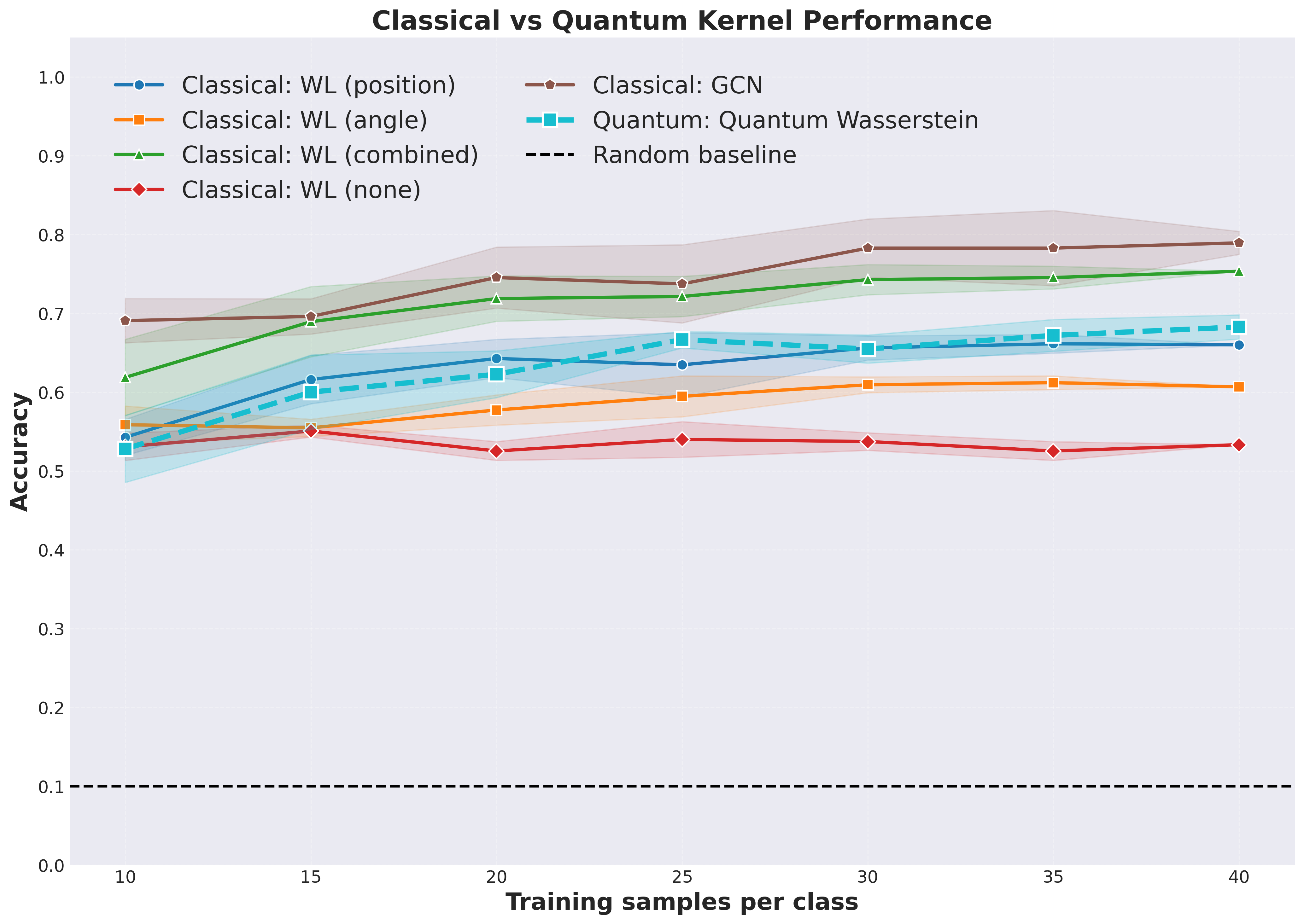}
    \caption{Comparison of our quantum kernel against classical baselines. The GCN achieves the highest accuracy, while our quantum kernel performs comparably to the positional WL variant and outperforms both angle-only and featureless WL kernels. Shaded regions indicate $\pm 1$ standard deviation over 5 seeds.}
    \label{fig:classical_vs_quantum}
\end{figure}

Fig.~\ref{fig:classical_vs_quantum} compares our quantum kernel (Wasserstein distance with sum aggregation) against classical baselines. The GCN achieves the highest mean accuracy, as expected. Among the WL kernels, the combined positional-angle variant performs best, slightly exceeding the mean accuracy of our quantum kernel. However, our quantum kernel matches the positional WL variant and yields a higher mean performance than both the angle-only and featureless WL kernels. This confirms the relevance of node-level features, with positional information playing a more critical role for this task.

\section{Related Work} 

Our work takes inspiration from Pasqal’s quantum evolution kernel (QEK) \cite{albrechtQuantumFeatureMaps2023, henryQuantumEvolutionKernel2021}, which uses neutral-atom (analog) quantum computers to naturally embed graph structures. QEK applies successive layers of mixing and natural evolution to encode graphs, then constructs distributions from Rydberg excitations. Djellabi et al. \cite{djellabiAttributedgraphsKernelImplementation2026} extended this to attributed graphs by encoding node features into local detuning fields, introducing the generalized-distance quantum-correlation kernel based on the local correlation matrix. Both approaches are based on two-level states and use only Rydberg lasers.  

In contrast, our approach introduces local Raman laser pulses which couple
the hyperfine ground states. This produces a three-level system for each atom, allowing for a richer Hilbert space, where node-features are encoded separately from the graph topology. We also use a different metric (Wasserstein), which allows us to obtain more accurate results compared to the Jensen-Shannon distance used in QEK.

\section{Conclusion}

In this paper, we proposed a quantum kernel method for node-attributed graphs embedded on neutral-atom quantum processors. We presented a simple prototypical circuit that encodes both structural and local feature information into the quantum evolution, and we showed how measurement distributions can be leveraged to define a kernel for classification. 

Our experimental results on the MNIST dataset demonstrated that the Wasserstein distance is the most effective metric among those tested, as its ability to bypass binning offers a practical advantage over the commonly used Jensen-Shannon distance. The quantum kernel achieved competitive performance with respect to selected classical graph kernels in a low-data regime, while the version without node-level features performed worse, confirming the importance of local information. The simplicity of our circuit, which uses a single encoding and executes the evolution layer for a very short time, did not allow us to surpass the mean performance of the WL combined classical kernel or the GCN. More refined designs with multiple, more elaborate encodings and longer evolution times could further improve performance, making this a promising direction to naturally encode node-attributed graphs. Interestingly, the optimizable weights $\omega_k$ of our kernel could be interpreted as a measure of relevance for the distributions, potentially offering insight into the choice of the graph's features that are most useful for the classification task. 
Regarding scalability, the encoding time on real quantum hardware for each graph scales linearly with the number, $O(|V|)$, of nodes, since preparing the atomic register and addressing atoms with local Raman pulses takes time proportional to the graph size. Estimating the fraction of global Rydberg excitations requires a constant number, $O(1/\epsilon^2)$, of shots, and computing correlations requires a logarithmic time $O(\log|V|/\epsilon^2)$ to maintain a target statistical error $\epsilon$. While executing this pipeline on real analog hardware will successfully eliminate the exponential bottleneck of classical state-vector simulation, demonstrating a strict computational or learning advantage over classical graph methods remains an open challenge. Furthermore, reducing and managing the classical $O(|V|^2)$ post-processing costs for the correlation matrix remains a critical focus for future large-scale implementations.

\section*{Acknowledgment}
We thank I. Sokolov and S. Acheche (PASQAL) for insightful discussions and support.






\bibliographystyle{IEEEtran}
\bibliography{references}

@article{aiQuantumGraphNeural2024,
  title={Towards {Quantum Graph Neural Networks}: An Ego-Graph Learning Approach},
  author={Ai, Xing and Zhang, Zhihong and Sun, Luzhe and Yan, Junchi and Hancock, Edwin R.},
  journal={arXiv:2201.05158},
  year={2024}
}

@article{albrechtQuantumFeatureMaps2023,
  title={Quantum Feature Maps for Graph Machine Learning on a Neutral Atom Quantum Processor},
  author={Albrecht, Boris and Dalyac, Constantin and Leclerc, Lucas and Ortiz-Guti{\'e}rrez, Luis and Thabet, Slimane and D'Arcangelo, Mauro and Cline, Julia R. K. and Elfving, Vincent E. and Lassabli{\`e}re, Lucas and Silv{\'e}rio, Henrique and Ximenez, Bruno and Henry, Louis-Paul and Signoles, Adrien and Henriet, Lo{\"\i}c},
  journal={Phys. Rev. A},
  volume={107},
  number={4},
  pages={042615},
  year={2023}
}

@article{Schuld_2019,
  title={Quantum Machine Learning in Feature Hilbert Spaces},
  author={Schuld, Maria and Killoran, Nathan},
  journal={Physical Review Letters},
  volume={122},
  number={4},
  pages={040504},
  year={2019}
}

@article{lecun2010mnist,
  title={MNIST handwritten digit database},
  author={LeCun, Yann and Cortes, Corinna and Burges, CJ},
  journal={ATT Labs [Online]. Available: http://yann.lecun.com/exdb/mnist},
  volume={2},
  year={2010}
}

@article{havlicekSupervisedLearningQuantumenhanced2019,
  title={Supervised learning with quantum-enhanced feature spaces},
  author={Havl{\'\i}{\v{c}}ek, Vojt{\v{e}}ch and C{\'o}rcoles, Antonio D and Temme, Kristan and Harrow, Aram W and Kandala, Abhinav and Chow, Jerry M and Gambetta, Jay M},
  journal={Nature},
  volume={567},
  number={7747},
  pages={209--212},
  year={2019}
}

@article{biamonteQuantumMachineLearning2017,
  title={Quantum Machine Learning},
  author={Biamonte, Jacob and Wittek, Peter and Pancotti, Nicola and Rebentrost, Patrick and Wiebe, Nathan and Lloyd, Seth},
  journal={Nature},
  volume={549},
  number={7671},
  pages={195--202},
  year={2017}
}

@article{djellabiAttributedgraphsKernelImplementation2026,
  title={Attributed-Graphs Kernel Implementation Using Local Detuning of Neutral-Atoms {Rydberg Hamiltonian}},
  author={Djellabi, Mehdi and Hecker, Matthias and Acheche, Shaheen},
  journal={arXiv:2509.09421},
  year={2026}
}

@article{ganCharactersGraphsInterpretable2023,
  title={Characters as graphs: Interpretable handwritten Chinese character recognition via Pyramid Graph Transformer},
  author={Gan, Ji and Chen, Yuyan and Hu, Bo and Leng, Jiaxu and Wang, Weiqiang and Gao, Xinbo},
  journal={Pattern Recognit.},
  volume={137},
  pages={109317},
  year={2023}
}

@inproceedings{ganStructureAwareHandwrittenTexta,
  title={Structure-Aware Handwritten Text Recognition via Graph-Enhanced Cross-Modal Mutual Learning},
  author={Gan, Ji and Zhou, Yupeng and Zhang, Yanming and Leng, Jiaxu and Gao, Xinbo},
  booktitle={Proc. IJCAI},
  pages={5154--5162},
  year={2025}
}

@book{hamiltonGraphRepresentationLearning,
  title={Graph Representation Learning},
  author={Hamilton, William L.},
  series={Synthesis Lectures on Artificial Intelligence and Machine Learning},
  publisher={Morgan \& Claypool Publishers},
  year={2020}
}

@article{henrietQuantumComputingNeutral2020a,
  title={Quantum Computing with Neutral Atoms},
  author={Henriet, Lo{\"\i}c and B{\'e}guin, Lucas and Signoles, Adrien and Lahaye, Thierry and Browaeys, Antoine and Reymond, Georges-Olivier and Jurczak, Christophe},
  journal={Quantum},
  volume={4},
  pages={327},
  year={2020}
}

@article{henryQuantumEvolutionKernel2021,
  title={Quantum Evolution Kernel: Machine Learning on Graphs with Programmable Arrays of Qubits},
  author={Henry, Louis-Paul and Thabet, Slimane and Dalyac, Constantin and Henriet, Lo{\"i}c},
  journal={Phys. Rev. A},
  volume={104},
  number={3},
  pages={032416},
  year={2021}
}

@inproceedings{kipfSemiSupervisedClassificationGraph2017a,
  title={Semi-Supervised Classification with Graph Convolutional Networks},
  author={Kipf, Thomas N. and Welling, Max},
  booktitle={Proc. Int. Conf. Learn. Representations (ICLR)},
  year={2017}
}

@article{kriegeSurveyGraphKernels2020,
  author    = {Kriege, Nils M. and Johansson, Fredrik D. and Morris, Christopher},
  title     = {A Survey on Graph Kernels},
  journal   = {Applied Network Science},
  volume    = {5},
  number    = {1},
  pages     = {6},
  year      = {2020},
  doi       = {10.1007/s41109-019-0195-3}
}

@book{maDeepLearningGraphs2021,
  title={Deep Learning on Graphs},
  author={Ma, Yao and Tang, Jiliang},
  publisher={Cambridge University Press},
  year={2021}
}

@article{fey2019fastgraphrepresentationlearning,
  title={Fast Graph Representation Learning with {PyTorch Geometric}},
  author={Fey, Matthias and Lenssen, Jan Eric},
  journal={arXiv:1903.02428},
  year={2019}
}

@misc{siglidis2020grakelgraphkernellibrary,
   author  = {Giannis Siglidis and Giannis Nikolentzos and Stratis Limnios and Christos Giatsidis and Konstantinos Skianis and Michalis Vazirgiannis},
  title   = {GraKeL: A Graph Kernel Library in Python},
  journal = {Journal of Machine Learning Research},
  year    = {2020},
  volume  = {21},
  number  = {54},
  pages   = {1--5},
}

@article{shervashidze2011weisfeiler,
  title = {Weisfeiler-Lehman Graph Kernels},
  journal = {Journal of Machine Learning Research},
  volume = {12},
  pages = {2539--2561},
  month = sep,
  year = {2011},
  author = {Shervashidze, N. and Schweitzer, P. and van Leeuwen, E. J. and Mehlhorn, K. and Borgwardt, K. M.}
}

@article{rubnerEarthMoversDistance,
  title={The Earth Mover's Distance as a Metric for Image Retrieval},
  author={Rubner, Yossi and Tomasi, Carlo and Guibas, Leonidas J.},
  journal={Int. J. Comput. Vis.},
  volume={40},
  number={2},
  pages={99--121},
  year={2000}
}

@article{zhangMotifEntropyGraph2023,
  title={Motif {Entropy Graph Kernel}},
  author={Zhang, Liang and Yi, Longqiang and Liu, Yu and Wang, Cheng and Zhou, Da},
  journal={Pattern Recognit.},
  volume={140},
  pages={109544},
  year={2023}
}

@article{givens1984wasserstein,
  title={A class of Wasserstein metrics for probability distributions},
  author={Givens, Clark R. and Shortt, Rae Michael},
  journal={Michigan Mathematical Journal},
  volume={31},
  number={2},
  pages={231--240},
  year={1984},
  publisher={University of Michigan, Department of Mathematics},
  doi={10.1307/mmj/1029003026}
}

@article{nguyenStreamingSlicedOptimal2026,
  title={Streaming {Sliced Optimal Transport}},
  author={Nguyen, Khai},
  journal={arXiv:2505.06835},
  year={2026}
}

@article{chenEnhancingGraphConvolutional2025,
  title={Enhancing Graph Convolutional Networks with an Efficient $k$-Hop Neighborhood Approach},
  author={Chen, Jiufang and Luo, Xin and Yuan, Ye and Wang, Zidong},
  journal={Inf. Fusion},
  volume={124},
  pages={103297},
  year={2025}
}

@article{tangQuantumGraphComputing2022,
  title={From {Quantum Graph Computing} to {Quantum Graph Learning}: A Survey},
  author={Tang, Yehui and Yan, Junchi and Hancock, Edwin R.},
  journal={arXiv:2202.09506},
  year={2022},
}

@article{yuQuantumGraphLearning2023,
        title={Quantum Graph Learning: Frontiers and Outlook}, 
      author={Shuo Yu and Ciyuan Peng and Yingbo Wang and Ahsan Shehzad and Feng Xia and Edwin R. Hancock},
      year={2023},
      journal={arXiv:2302.00892},
      eprint={2302.00892},
      archivePrefix={arXiv},
      primaryClass={cs.LG},
}

@article{chenHybridQuantumClassicalGraph2021,
  title={Hybrid {Quantum-Classical Graph Convolutional Network}},
  author={Chen, Samuel Yen-Chi and Wei, Tzu-Chieh and Zhang, Chao and Yu, Haiwang and Yoo, Shinjae},
  journal={arXiv:2101.06189},
  year={2021}
}

@article{Silv_rio_2022,
  title={Pulser: An open-source package for the design of pulse sequences in programmable neutral-atom arrays},
  author={Silv{\'e}rio, Henrique and Grijalva, Sebasti{\'a}n and Dalyac, Constantin and Leclerc, Lucas and Karalekas, Peter J. and Shammah, Nathan and Beji, Mourad and Henry, Louis-Paul and Henriet, Lo{\"i}c},
  journal={Quantum},
  volume={6},
  pages={629},
  year={2022}
}

@article{morrisWeisfeilerLemanGo,
  title={{Weisfeiler} and {Leman} go sparse: Towards scalable higher-order graph embeddings},
  author={Morris, Christopher and Rattan, Gaurav and Mutzel, Petra},
  journal={arXiv:1904.01543},
  year={2020}
}

\end{document}